\documentclass[apj,twocolumn]{emulateapj}

\usepackage{graphicx}
\newcommand{\gapprox}{\,\rlap{\lower 2.5pt 
\hbox{$\sim$}}\raise 1.5pt\hbox{$>$}\,}
\newcommand{\lapprox}{\,\rlap{\lower 2.5pt 
\hbox{$\sim$}}\raise 1.5pt\hbox{$<$}\,}

\newcommand{\zh}{[{{\rm Z}/{\rm H}}]}

\slugcomment{Submitted to ApJ Letters}
\shortauthors{Maraston et al.}
\begin{document}
\title{Modeling the color evolution of luminous red galaxies - improvements with empirical stellar spectra}
\author{
         C. Maraston\altaffilmark{1},
        G. Str\"omb\"ack\altaffilmark{1},
	 D. Thomas\altaffilmark{1},
	 D.A. Wake\altaffilmark{2},
	 R.C. Nichol\altaffilmark{1}
}
\altaffiltext{1}{Institute of Cosmology and Gravitation, Mercantile House, Hampshire Terrace, Portsmouth, PO1 2EG, UK} \email{claudia.maraston@port.ac.uk}
\altaffiltext{2}{Department of Physics, University of Durham, South Road, Durham, DH1 3LE, UK}
\begin{abstract}
Predicting the colors of Luminous Red Galaxies (LRGs) in the Sloan
Digital Sky Survey (SDSS) has been a long-standing problem. The $g,r,i$ colors of LRGs are inconsistent with stellar population models over the redshift range $0.1<z<0.7$. The $g-r$ colors in the models are on
average redder than the data (of the order 0.1 mag) while the
$r-i$ colors in the models are bluer (by 0.05 mag) towards low redshift. Beyond redshift
0.4, the predicted $r-i$ color becomes instead too red, while the predicted $g-r$ agrees with the data.
We provide a solution to this problem, through a combination of new astrophysics and a fundamental change to the stellar population modeling. We find that the use of the empirical library of Pickles (1998), in place of theoretical libraries based on model atmosphere calculations, modifies the evolutionary population synthesis predicted colors exactly in the way suggested by the data, i.e., gives a redder $r-i$ color, and a bluer $g-r$ color, in the observed frame at $z=0.1$. The reason is a lower flux in the empirical libraries, with respect to the theoretical ones, in the wavelength range $5500-6500$~\AA. The discrepancy increases with decreasing effective temperature independently of gravity. This result has general implications for a variety of studies from globular clusters to high-redshift galaxies. The astrophysical part of our solution regards the composition of the stellar populations of these massive Luminous Red Galaxies. We find that on top of the previous effect one needs to consider a model in which $\sim~3\%$ of the stellar mass is in old metal-poor stars. Other solutions such as substantial blue Horizontal Branch at high metallicity or young stellar populations can be ruled out by the data.
Our new model provides a better fit to the $g-r$ and $r-i$ colors of LRGs and gives new insight into the formation histories of these most massive galaxies. Our model will also improve the {\it k}- and evolutionary
corrections for LRGs which are critical for fully exploiting present and future galaxy surveys.
\end{abstract}
\section{Introduction}
Age-dating the stellar populations of galaxies provides astronomers
with a cosmic timescale which, being ruled by stellar evolution, is
independent of cosmological models and allows the use of galaxies as
cosmological probes \citep[e.g.][]{jimloe02}. The interpretation of
galaxy spectra in terms of stellar populations is also the only
effective way of reconstructing their star-formation histories, which
allows one to get clues on the still poorly-known process of galaxy
formation. There is a long history of research in this area. We provide here a selection of recent works \citep{kauetal03,thoetal05,neletal05,beretal06,panetal07,jimetal07} and refer the reader to Renzini (2006) for a comprehensive review of studies at both low and high redshift.

Studies of galaxy evolution based on the Sloan Digital Sky Survey (SDSS) data of Luminous (massive) Red Galaxies (LRGs) have highlighted a potentially fundamental problem with the rest-frame optical of stellar population models. 
As first noticed by \citet{eisetal01}, the models are systematically too red in the $g-r$~observed-frame in the redshift range 0.1 to 0.4 (see Figure~1, left-hand panel). This discrepancy could not be cured by adopting more complex stellar population models with various star formation histories, different stellar population model codes or empirical galaxy spectra. To be able to analyse the data, \cite{eisetal01} applied a shift of 0.08 mag to the models.

\cite[][W06]{waketal06} extended this study by including the $r-i$ color as a further constraint and by adding the 2dF SDSS LRG and Quasar (2SLAQ) survey sample \citep{canetal06} to increase the redshift range to $z\sim 0.8$, demonstrating further discrepancies between the models and data. While the models are too red in the observed frame $g-r$ at low redshift, the $r-i$ color turns out to be too blue (see Figure~1, left-hand panel). This pattern changes with increasing redshift such that the predicted $r-i$ color becomes too red, while the predicted $g-r$ matches the data (all observed frame). Figure~1 in W06 demonstrates that the addition of star formation does not solve the problem. Indeed, the light of young, blue stars while curing the $g-r$ syndrome at the lowest redshifts, worsens the discrepancy in the $r-i$ color at low redshifts and in the $g-r$ color at high redshifts, such that colors in the model become too blue. The origin of this mismatch has remained a puzzle so far, and has serious consequences for the interpretation of current observations and the planning of future galaxy surveys. In this paper we provide a solution to this problem and present a working stellar population model. 

The paper is arranged as follows. In Section 2 we recall the features
of the luminous red galaxy sample. In Section 3 we present the working solution and we 
comment on other discarded options. A conclusive discussion is placed in Section 4.
\begin{figure*}
\centering\includegraphics[width=0.75\textwidth]{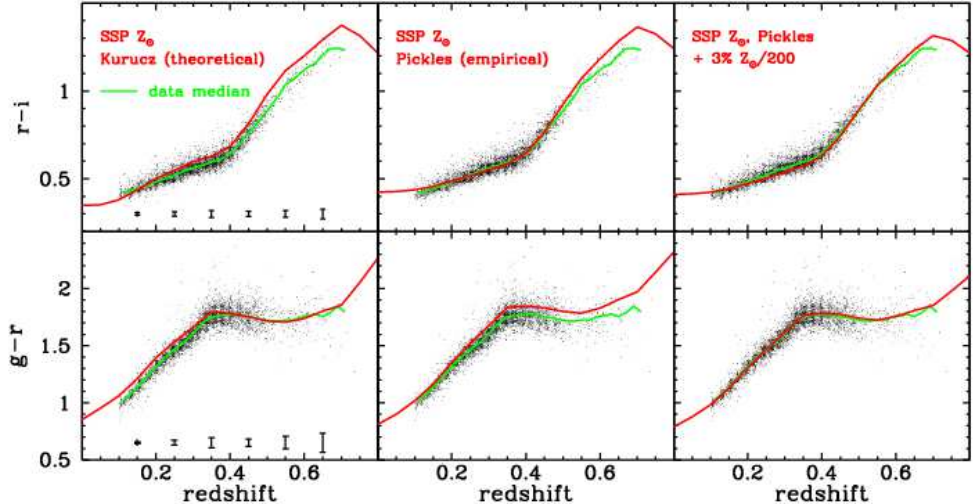}
\caption{The $g-r$ and $r-i$ colors of LRGs as functions of redshift (points; data from W06). The median is given by the green line. Typical errors as function of redshift indicated by the error bars. {\it Left panels}.  A solar-metallicity passively evolving,
single-burst model with an age of 12 Gyr at redshift zero (red line). {\it Middle panels}. Same data as in the left-hand panel. The stellar population model uses the Pickles (1998) empirical spectral library instead of the theoretical one (see text). {\it Right panels}. Same data as in the left-hand panel with a composite model with $3\%$ by mass of metal-poor stars. Both the metal-rich and the metal-poor component are 12 Gyr old at redshift zero. The metal-rich component uses the Pickles (1998) empirical spectral library. 
\label{colors}}
\end{figure*}
\section{The data sample}
W06 extract their sample of LRGs from the SDSS multi-color photometry \citep[SDSS,][]{yoretal00}. Spectroscopic redshifts are taken from the SDSS in the redshift range $0.15 < z < 0.37$, and from the 2SLAQ survey \citep{canetal06} in the redshift range $0.45 < z < 0.8$. At $z>0.4$ we use the deeper multi-epoch data in stripe 82. LRGs from
both surveys were selected in a consistent way using the same
rest-frame color cuts, under the assumption that they were old,
passively evolving galaxies, in order to make sure that the same
galaxy population is sampled at all redshifts (see W06 for details).

Obviously, the sample definition depends on the color selection,
which in turn depends on the model adopted to identify the LRGs. This
is what we should call model-dependent data setting. One of the worse
consequences of the discrepancy between data and models is that in order to minimise the effect of incorrectly $k$-correcting the data a conspicuous part of the initial data base had to be discarded.
This fraction was as substantial as 85\% in W06. Hence, besides the general
astrophysical goal of understanding the stellar populations of
luminous, massive galaxies, there is the more practical need to
optimally exploit the precious data emerging from present and
future surveys of LRGs to high redshift.

To ensure self-consistency, the data used in this paper have been re-selected using 
the \cite[][M05]{mar05} models.

In the following we assume that galaxies start forming
stars at a redshift of five, as suggested by the local fossils
\citep[e.g.][]{kauetal03,neletal05,beretal06,jimetal07,thoetal05},
which fixes their present age to be about 12 Gyr\footnote{Wake et
al. used 13 Gyr, but this small difference does not impact on the
overall conclusions.}.
\section{The working model}
\begin{figure}
\includegraphics[width=\linewidth]{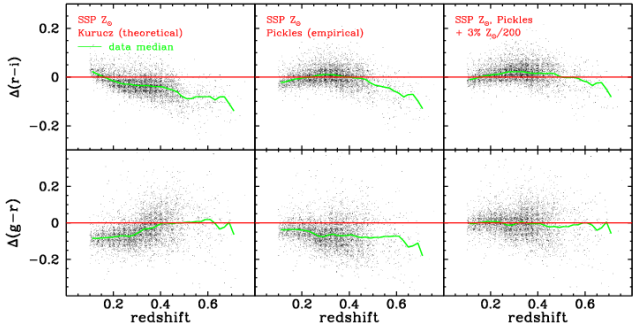}
\caption{Residuals of the plots shown in Figure~\ref{colors}.\label{residuals}}
\end{figure}

The problem as explained in Section~1 is illustrated in the left-hand panel of Figure~\ref{colors}, which is a remake of Figure~1 by W06. The red line refers to a single-burst model (Simple Stellar Population, SSP) with solar metallicity and an age of 12 Gyr at redshift zero. Residuals are shown in Figure~\ref{residuals}. The right-hand panels of Figures~\ref{colors} and~\ref{residuals} show the solution to the problem using our new model. The latter is based on two modifications; the replacement of theoretical stellar atmospheres by empirical ones, and the inclusion of a small (3\% by mass), metal-poor subcomponent. Details of this model are described in the following sections.
\begin{figure*}
\centering\includegraphics[width=0.6\linewidth]{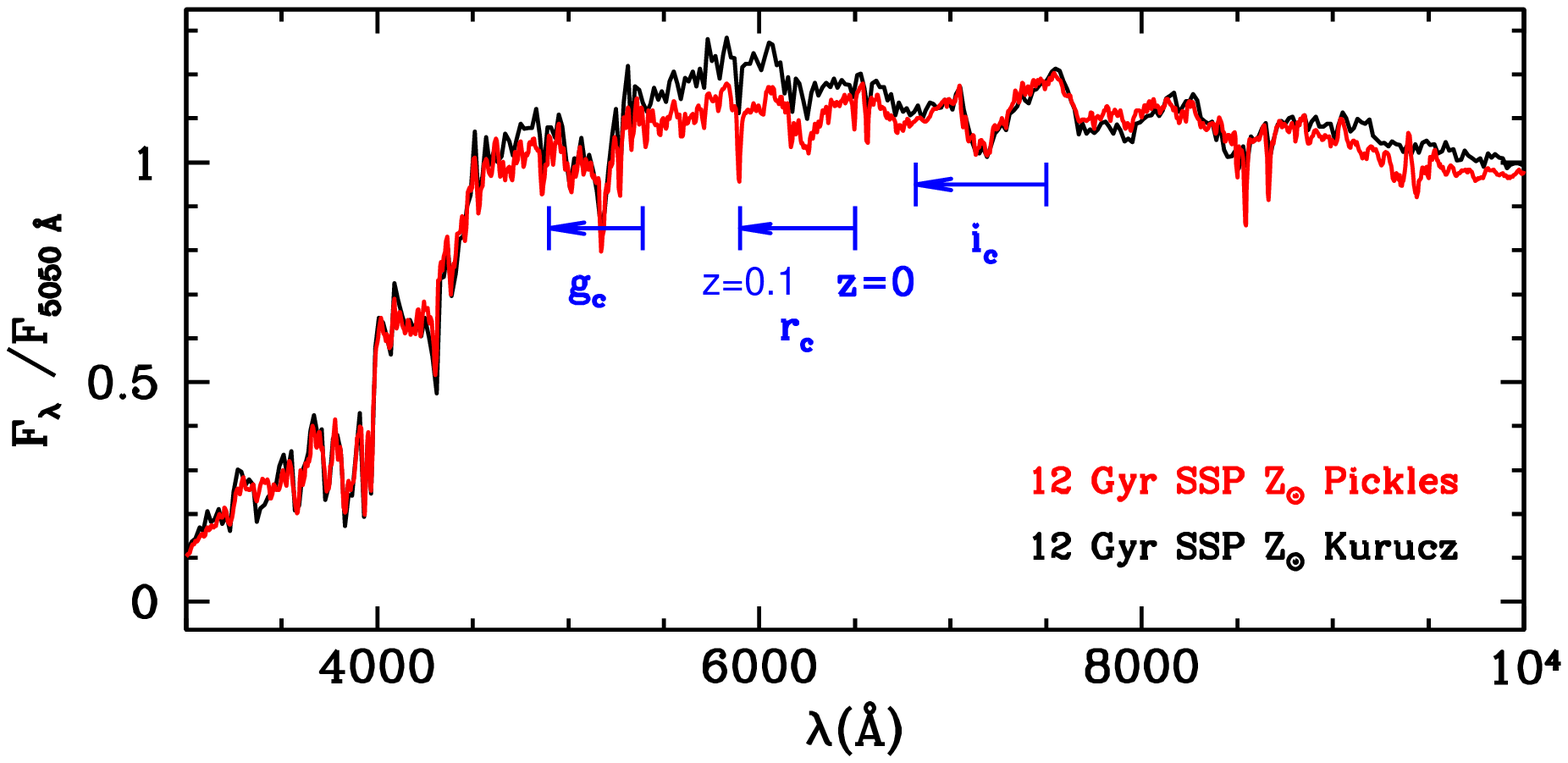}
\caption{The effect of the input spectral library. The spectral energy distributions of 12-Gyr-old, solar-metallicity
SSPs using theoretical (black) and empirical (red, from Pickles~1998) spectral library. Note the flux excess in
the theoretical spectrum between $\sim5500$ and $\sim6500$~\AA. Also shown are the central wavelengths of 
the $g$, $r$, $i$ SDSS filters at redshift zero and 0.1.
\label{atmo_effect}}
\end{figure*}

\subsection{Inclusion of empirical stellar atmospheres}
The simultaneous mismatch in opposite directions of colors sampling close rest-frame wavelengths 
was suggestive to us of a shortcoming in the stellar spectra that are used in the population synthesis models \citep[e.g.][based on the Kurucz~1979 library]{lejetal97}. To explore this path we consider empirical spectral libraries as a substitute to the theoretical ones and compute stellar population models using such libraries as input to the M05 code.

Several high-to-intermediate resolution libraries of flux calibrated empirical 
stellar spectra, intended for use in evolutionary population synthesis, have 
been constructed over the last decade, e.g. \cite{pic98}, STELIB \citep{leboretal03}, MILES \citep{sanetal06} and ELODIE.v3.1 \citep{pruetal07}. For the aim of this work, which is to study the redshift evolution of spectra, the Pickles library is the only suitable one, due to its wide wavelength coverage ($1150-25000\;$\AA). Models constructed from the other empirical libraries were also considered, but their wavelength coverage is too narrow (3900-6800~\AA~for the ELODIE.v3.1. and 3500-7500~\AA~for MILES, respectively) to make them suitable for studying the evolution of spectra with redshift.

The mapping of stellar atmospheric parameters in the Pickles library is very good, especially the lower 
main sequence and the tip of the red giant branch. The rather low spectral resolution of the Pickles library 
($R\,\approx\,500$) is not an issue here, since we are only interested in broadband colors. However, the library does not contain enough stars to cover all important evolutionary phases for non-solar metallicities. Therefore, this exercise focuses on models with solar metallicity, the most relevant for massive galaxies such as LRGs.

The effect of the input stellar library is shown in Figure~\ref{atmo_effect}, which displays the spectral energy distributions of two 12 Gyr, solar metallicity SSPs models that only differ in their input spectral library. The red line shows the one based on the P98 empirical library, whereas the black one is based on the \citet{lejetal97} theoretical library. A flux excess is evident in the theoretical spectrum between $\sim~5500$ and $\sim~6500$~\AA. This discrepancy increases with decreasing effective temperature independently of gravity, and points to problems in the model atmosphere calculations at temperatures below $\sim 6000\;$K.

Also shown in the figure are the central wavelengths of the $g$, $r$, $i$ SDSS filters at redshifts zero and 0.1. The lower flux in the empirical spectrum is sampled by the $r$-filter at redshift 0.1, which explains the redder $r-i$ colors (by $\sim 0.06$ mag) of the empirically-based models with respect to the theoretical ones. For the same reason, the $g-r$ at $z=0.1$ is bluer by $\sim 0.05$ mag. The comparison with SDSS data is shown by the middle panels of Figures~\ref{colors} and~\ref{residuals}. The inclusion of the empirical library rectifies the position of the models with respect to the data at the lowest sampled redshift. The predicted $g-r$ color becomes bluer and the $r-i$ color redder around $z\sim 0.1$. The trend is the opposite at higher redshifts; the observed-frame $g-r$ color gets redder and the $r-i$ color slightly bluer around $z\sim 0.6$. Hence the model has improved mainly at lower redshifts, and most significantly in the $r-i$ color. Further modifications to the model are still necessary to track correctly the evolution with redshift. However, the situation has become much clearer with the inclusion of empirical atmospheres; the model spectral energy distributions (SEDs) need to become systematically bluer shortward of the wavelength range sampled by $r-i$ at $z\sim 0.1$.

\subsection{The metal-poor subcomponent}
As the middle panel of Figure~\ref{colors} shows, the match between models and data still requires a slightly bluer $g-r$ colors at low redshift, but now both bluer $g-r$ and $r-i$ colors at high redshift. The flatness of $g-r$ beyond $z=0.4$ further suggests that the required bluening component must be slowly evolving with look-back time. This disfavors the option of adding a component of residual star formation (see W06 and Section~3.3). A metal-poor old sub-component is the best candidate.

Metal-poor stellar populations have blue turnoff's and very often Blue
Horizontal Branches (BHBs). A BHB from a metal-poor stellar population
arises due to the relatively high effective temperature of
the evolving stars, even without assuming large amount of mass-loss
during the Red Giant Branch phase. Calculations show that a
stellar population with metallicity $\zh\sim -2.2$ has already
developed a BHB at ages of $\sim 6 $ Gyr with a mass-loss efficiency
consistent with calibrations from Milky Way globular clusters (Figure~11 in M05.) 

The difference in spectral energy distributions between a solar-metallicity model and the same contaminated by metal-poor stars (3\% in mass) is such that the composite SED has more flux in the optical bands, while leaving unaltered the flux longward $\sim~9000$~\AA\ \citep{martho00}. This helps the $g-r$ color to become bluer without perturbing the $r-i$ color at the lowest redshifts. At the highest redshifts it makes both colors bluer as required by the data.

As mention above, components of metallicities other than solar cannot be based on the empirical spectral library of Pickles. The metal-poor subcomponent is therefore taken from the original M05 model based on theoretical stellar atmospheres. Even if the inadequacies of the theoretical atmosphere in the optical wavelength region at solar metallicities (see Figure~\ref{atmo_effect}) propagates to the lowest metallicities, we can safely neglect this effect, as the metal-poor subcomponent contributes only 3\% to the total mass of the composite stellar population. 

The final composite solution and the color residuals are shown in the right-hand panels of Figures~\ref{colors} and \ref{residuals}. The redshift evolution of the SDSS data are now well matched by the model in both the $g-r$ and $r-i$ colors with a residual discrepancy of only 0.02~mag in $r-i$ around $z\sim 0.3$.

\subsection{Discarded options}
The inclusion of a BHB in the metal-rich component is less attractive, as a very large fraction (50\%) of the population would need to develop the BHB.

We also tested the possibility of including recent star formation on top of the passively evolving SSP based on the Pickles stellar library. We included a small subcomponent with constant star formation rate since $z=5$ summing up to a mass fraction of 3\% at $z=0$. We confirm that the blueing of observed-frame $g-r$ becomes too large around $z=0.6$ with respect to $z=0.1$. This effect could be avoided only by assuming that the level of residual star formation in LRGs has been increasing steadily at least since redshift $z=0.7$, which seems contrived. 

Finally, the LRGs in the SDSS are known to host $\alpha$/Fe-enhanced stellar populations \citep{eisetal03}. We tested the effect of including such enhancements by calculating models using the $\alpha/Fe$ enhanced tracks of \cite{saletal00}. This yields a slight bluening of both $g-r$ and $r-i$ only at the lowest redshifts, while both colors at $z>0.4$ remain unchanged. This is because $\alpha$/Fe enhancement impacts on the tracks only at old ages \citep{saletal00}. Hence this option does not represent a viable solution.

\section{Discussion}
In this paper we provide a working stellar population model for luminous red galaxies (LRG), which provides a substantial improvement to a long-standing mismatch with the data from the SDSS survey \citep{eisetal01,waketal06}.

The new model adopts empirical stellar spectra from the library of Pickles~(1998) in place of the theoretical ones at solar metallicity in the evolutionary population synthesis. We found that an excess flux around $6000$~\AA\ rest-frame in the theoretical 
spectra was responsible for making the synthetic $r-i$~color too blue at redshift 0.1, and the $g-r$~ color too red. This finding impacts on a variety of studies which involve the $B-V$ and $V-R$ rest-frame and might be the cause for a mismatch between stellar population models and Milky Way Globular Cluster data in $B-V$ as highlighted in M05.

The new model further includes the addition of a small (3\% by mass), metal-poor subcomponent, 
being coeval to the old and metal-rich dominant population. The metal-poor old subcomponent can better match the observed color and color evolution because of its slow evolution with redshift in comparison to a young sub-population, whose color evolution is much stronger. This model was constrained using data in the redshift range 0.1 to $\sim~0.8$. An early version of this model was used in Cool et al.~(2008) for studying the luminosity function up to a redshift 0.8. Though the model allowed for a better match towards the highest redshifts, the synthetic $g-r$ was still too red at the lowest redshifts (cfr. Figure~13 in Cool et al. 2008), and the $r-i$ was not well recovered. This early model did not include the empirical spectra. For this reason a larger metallicity for the dominant population needed to be assumed, and as a consequence a higher fraction by mass of metal-poor stars. The use of the empirical spectra has solved the remaining problems.

For the discussion of the possible origins of a metal-poor subcomponent in massive galaxies we refer to \cite{martho00}. It needs to be assessed whether the requirement that 3$\%$~by mass of stars in LRGs are very metal-poor is just the reflection of the metallicity gradient known to be present in massive galaxies or is indeed caused by the presence of a very metal-poor subcomponent. The latter could originate from accretion of metal-poor dwarf satellites during the evolution of the galaxy. Future investigations on this question will be very valuable.

The models are available at www.maraston.eu.
\acknowledgments
We acknowledge stimulating discussion with Jim Gunn and Andrew Pickles. 
CM holds a Marie Curie Excellence Team Grant MEXT-CT-2006-042754
of the Training and Mobility of Researchers programme financed by the
European Community.
\end{document}